\documentclass[runningheads]{llncs}
\usepackage[T1]{fontenc}
\usepackage{graphicx,verbatim}
\usepackage{amsmath}
\usepackage{amssymb}
\usepackage{xcolor}

\usepackage[colorlinks=true,
            linkcolor=blue,
            citecolor=blue,
            urlcolor=blue]{hyperref}

\usepackage{booktabs}
\usepackage{siunitx}
\usepackage{multirow}
\usepackage{tabularx}

\begin{document}
\title{Measuring Prediction Uncertainty in Neural Cellular Automata}
%

\author{Ario Sadafi\inst{1,2,3,4} \and Michael Deutges\inst{1} \and
Nassir Navab \inst{3,4} \and \\
Carsten Marr \inst{1,2,4,5,6,7}
\thanks{\raggedright Correspondence: \{carsten.marr, ario.sadafi\}@helmholtz-munich.de}}
\renewcommand\footnotemark{}

\authorrunning{A. Sadafi et al.}

\institute{Computational Health Center, Helmholtz Munich, Neuherberg, Germany \and 
Helmholtz AI, Helmholtz Munich, Neuherberg, Germany \and
Computer Aided Medical Procedures, Technical University of Munich, Munich, Germany \and Munich Center for Machine Learning, Munich, Germany \and Department of Medicine III, Ludwig-Maximilian-University Hospital, Munich, Germany \and Department of Physics, University of Munich, Munich, Germany \and German Cancer Consortium (DKTK), partner site Munich, Germany}

\maketitle              
\begin{abstract}
Neural cellular automata (NCA) provide a lightweight alternative to encoder–decoder segmentation networks.
However, it can be difficult to decide when a prediction should be trusted. 
Here, we study uncertainty estimation for NCA-based medical image segmentation without modifying the underlying architecture or retraining the model. Our approach is motivated by viewing the NCA as a dynamical system where convergent attractors correspond to confident predictions.

Concretely, we propose \textit{resilience}, a simple measure that leverages the intrinsic iterative structure of NCAs by probing the stability of the final prediction under small perturbations of the automaton state. Predictions that return to the same solution are deemed confident, while those that change substantially are flagged as uncertain.

We evaluate uncertainty by its ability to predict segmentation quality using selective prediction metrics ($\Delta$Dice@90 and AURC) and ranking metrics (AUROC and AUPRC). Across multiple medical segmentation benchmarks, \textit{resilience} identifies failure cases more reliably than baselines, improving trust and safety in NCA-based models.


\end{abstract}
\section{Introduction}
Medical image segmentation is a foundational step in many clinical and biomedical workflows, from quantifying structures in microscopy to delineating lesions and organs in radiology and endoscopy \cite{wang2022medical}. Yet even strong segmentation models can fail silently on unusual anatomy, poor image quality, or domain shifts across devices and sites. In practice, this makes reliability as important as accuracy. A usable system should not only output a mask, but also provide a meaningful signal of when that mask can be trusted, reviewed, or rejected \cite{gonzalez2022distance}.

Uncertainty estimation has become the standard route to reliability, enabling selective prediction, quality control, and human-in-the-loop triage \cite{dolezal2022uncertainty}. Most existing approaches, however, are developed for feed-forward architectures that produce predictions in a single pass, where uncertainty is tied to predictive probabilities, ensembling, or test-time augmentation. While effective in many settings, these techniques do not explicitly account for models whose predictions emerge from an iterative process and may therefore miss uncertainty cues that arise from the dynamics of inference itself.

Neural Cellular Automata (NCA) offer a lightweight and conceptually appealing alternative for segmentation \cite{mordvintsev2020growing}. Rather than relying on an encoder– decoder backbone, an NCA maintains a spatial state and repeatedly applies a shared local update rule until a segmentation is formed \cite{deutges2025neural,kalkhof2023med,kalkhof2023m3d,lemke2025octreenca}. This dynamical view changes the nature of uncertainty. For an iterative model, an apparently sharp final mask can still be unreliable if the NCA dynamics have not converged to a stable prediction, i.e. if it oscillates between competing solutions, or if small disturbances steer the system toward a different outcome. These behaviors are described in terms of stability and attractors—properties that are not captured by a single forward-pass confidence score. 
Furthermore, NCA models have been applied beyond segmentation to tasks such as image classification and generative modeling~\cite{deutges2024neural,mordvintsev2020growing,yang2025attention,yang2025hierarchical}. This broader applicability motivates uncertainty measures that operate directly on the underlying dynamical system rather than on a task-specific architectures.

This work proposes a simple, NCA-native uncertainty measure that we call \textit{resilience}. The central idea is to probe whether the final segmentation behaves like a stable solution of the learned dynamics. Starting from the final NCA state, a small perturbation is injected into the latent state and the automaton is allowed to evolve for a few additional steps. If it recovers to the same segmentation, the prediction is considered confident; if the prediction changes substantially, the output is considered uncertain. Unlike Monte Carlo dropout \cite{gal2016dropout}, which samples variability through stochastic subnetworks and requires architectural modification to include dropout layers, \textit{resilience} tests the stability of the learned dynamical system and can be applied without retraining.

To place \textit{resilience} in context, we evaluate several complementary uncertainty signals as baselines under a common protocol that assigns one scalar uncertainty score per image and measures how effectively uncertainty ranks predictions. Evaluation is performed across diverse medical segmentation benchmarks spanning microscopy, histology, dermoscopy, and endoscopy. Uncertainty is quantified using selective prediction criteria and failure-detection metrics enabling a consistent comparison across datasets with different baseline accuracies.

The main contributions are: (1) We provide, to our knowledge, the first systematic study of uncertainty estimation for NCA-based medical segmentation, covering four imaging modalities and explicit distribution shifts under a unified evaluation protocol. (2) We introduce \textit{resilience}, a training-free perturb-and-recover uncertainty measure that directly probes the stability of the NCA’s learned dynamics, requiring neither architectural changes nor retraining. We test our approach on five benchmarks and show that \textit{resilience} yields a strong and consistent signal for uncertainty-aware deployment of NCA-based models.

Our implementation, including all uncertainty baselines and evaluation scripts is publicly available at \url{https://github.com/marrlab/resilience}.

\section{Methods}

\subsection{Neural Cellular Automata for Segmentation}
We model image segmentation using Neural Cellular Automata (NCA), which defines a discrete dynamical system over a latent state \cite{mordvintsev2020growing,kalkhof2025med}.
Given an RGB input image $x \in \mathbb{R}^{H \times W \times 3}$, the NCA maintains a state tensor
\[
S_t \in \mathbb{R}^{H \times W \times C}, \quad t = 0,1,\dots,T,
\]
where $C$ denotes the number of state channels.
The initial state $S_0$ is constructed from the input image by concatenation with zero-initialized latent channels.

The states evolve according to a learned update rule
$S_{t+1} = F_\theta(S_t),$
where $F_\theta$ is a local neural network applied identically at every spatial location and time step.
After $T$ iterations, a segmentation prediction is obtained from a designated output channel
$m = o(S_T),$
where we denote $o(\cdot)$ as readout operator, and $m \in \{0,1\}^{H \times W}$ as the predicted binary mask.
Crucially, the NCA defines an explicit trajectory $\{S_t\}$ whose convergence behavior is governed by the learned dynamics.

\subsection{Resilience Uncertainty via Perturb-and-Recover}

\begin{figure}[t]
\centering
\includegraphics[width=\textwidth,page=1,trim=1cm 7.5cm 3cm 1.5cm,clip]{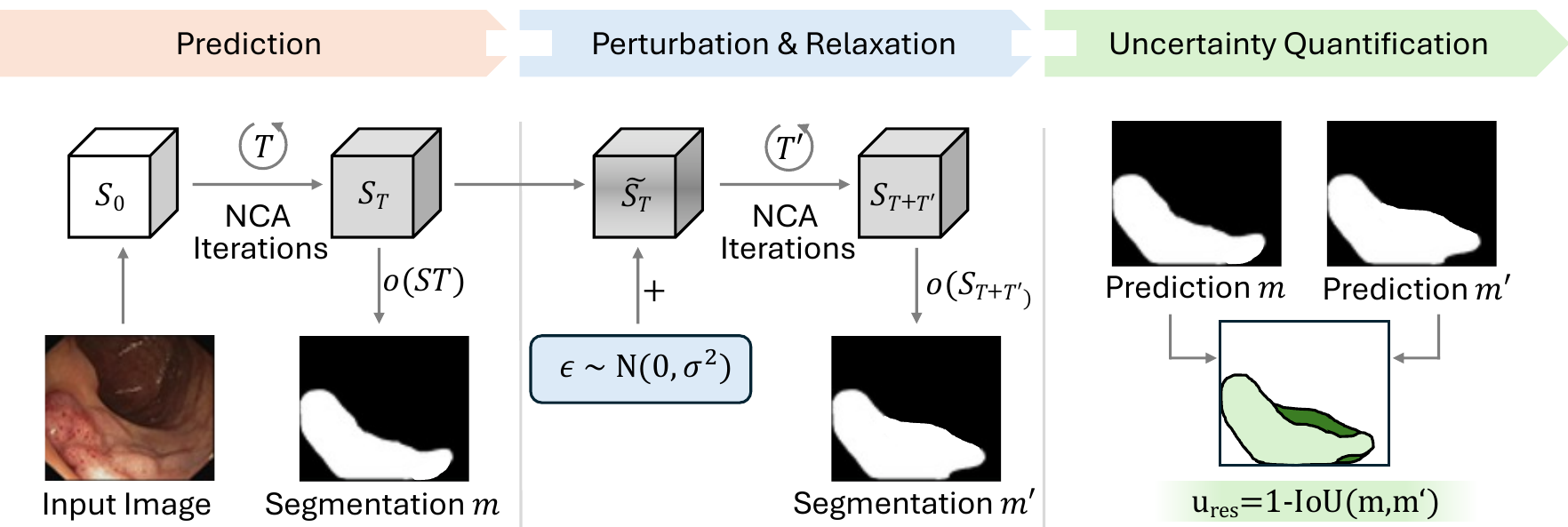}
\caption{Uncertainty quantification via perturb-and-recover in Neural Cellular Automata (NCA).
After iterating the NCA for $T$ steps on an input image, a final state $S_T$ and segmentation mask $m$ are obtained.
To probe the stability of the learned dynamics, Gaussian noise $\epsilon$ is injected and the system is allowed to relax for an additional $T'$ steps.
The resulting prediction $m'$ is compared to the original mask using the Intersection-over-Union (IoU).
The \textit{resilience} method $u_{\mathrm{res}} = 1 - \mathrm{IoU}(m, m')$ measures sensitivity to state perturbations:
low values indicate convergence to a stable attractor, while high values indicate unstable (and uncertain) segmentations.}
\label{fig1}
\end{figure}

We exploit the dynamical nature of the NCA to estimate prediction uncertainty by probing the stability of the final state under perturbations (Figure~\ref{fig1}).

For a given test image, we first run the NCA for $T$ steps to obtain the final state $S_T$ and its associated prediction $m$.
We then perturb the final state by adding Gaussian noise,
\[
\widetilde{S}_T = S_T + \varepsilon, \qquad \varepsilon \sim \mathcal{N}(0, \sigma^2 I),
\]
where $\sigma > 0$ controls the perturbation magnitude and $I$ denotes the identity over all state dimensions. Thus, a separate Gaussian noise value is sampled for every element of the state tensor, i.e., for every pixel location and every state channel; the same noise value is not shared across channels.
Importantly, the input image remains fixed; only the latent NCA state is perturbed. 

Starting from the perturbed state $\widetilde{S}_T$, we apply an additional $T'$ NCA update steps,
\[
S_{T+T'} = F_\theta^{T'}(\widetilde{S}_T),
\]
and compute a segmentation prediction after recovery as
$m' = o(S_{T+T'})$.

We quantify the degree of recovery by measuring the Intersection over Union $\text{IoU}(m,m')=\frac{|m \cap m'|}{|m \cup m'|}$ between the original and relaxed predictions. 
We define the uncertainty measure \textit{resilience} as
\[
u_{\mathrm{res}} = 1 - \text{IoU}(m,m').
\]

In all experiments, we used a perturbation magnitude of $\sigma = 0.02$ and a relaxation length of $T' = 12$. \textit{Resilience} was computed using one perturb-and-recover rollout per image.

Low values of $u_{\mathrm{res}}$ indicate that the NCA dynamics recover to an equivalent segmentation after perturbation, suggesting a stable attractor (see Figure \ref{fig1}).
Conversely, high values indicate sensitivity to small state perturbations and correspond to unstable predictions.

Our proposed uncertainty measure, \textit{resilience}, can be interpreted as a local stability probe of the learned dynamical system around the final state $S_T$.
When $\widetilde{S}_T$ lies within the basin of attraction of a fixed point, perturbations return back to $S_T$ under further NCA iterations and $m' \approx m$.
In contrast, for challenging samples, the dynamics may result in divergent predictions and therefore increased uncertainty.

\section{Experiments and Results}
\subsection{Datasets}
We evaluate \textit{resilience} on five medical image segmentation benchmarks. All tasks are formulated as binary segmentation.

\begin{itemize}
\item ClinicDB \cite{bernal2015wm}: ClinicDB is a colonoscopy benchmark consisting of endoscopic frames with corresponding pixel-level polyp masks. We use a deterministic split to obtain 428 training, 91 validation, and 93 test images. Masks are binary polyp silhouettes.

\item DSB 2018 \cite{caicedo2019nucleus}: We utilize the 2018 Data Science Bowl benchmark for nuclei segmentation. We generate a deterministic split by sorting the 671 available case folders and partitioning them 80/10/10, resulting in 537 training, 67 validation, and 67 test images. Instance masks are unified into a single binary foreground mask per image.

\item ISIC 2017 \cite{codella2018skin}: This dataset for skin lesion segmentation is used with its official splits: 2,000 training, 150 validation, and 600 test images. 

\item Kvasir-SEG \cite{jha2019kvasir}: Consisting of 1,000 colonoscopy images, we define this task as a polyp-vs-background segmentation. We apply a 70/15/15 split resulting in 700 train, 150 validation, and 150 test images.

\item NuInsSeg \cite{mahbod2024nuinsseg}: This benchmark spans multiple histology tissue domains for nuclei segmentation. We split the dataset into 465 training, 99 validation, and 101 test images, ensuring no slide-level overlap across splits.

\end{itemize}
To assess uncertainty under distribution shifts and harsher scenarios, we augmented the test set for ClinicDB and Kvasir-SEG datasets by applying geometric and photometric perturbations to the original test samples

\subsection{Training}
The same training settings across all datasets are used without dataset-specific tuning. While this is not optimal for maximizing segmentation performance on each benchmark, it provides a consistent and sufficiently challenging regime in which prediction errors, and therefore uncertainty, remain observable and comparable across datasets.
We train an NCA segmentation model with C=64 state channels, fire rate 0.5, hidden size 128, and three RGB input channels. The fire rate denotes the
probability that a cell updates its state at each NCA step. All models are optimized with Adam (learning rate $10^{-3}$, weight decay $10^{-2}$) for 50 epochs using a batch size of 2. Training uses a rollout-length schedule: at each iteration we sample the number of NCA update steps $T \sim \mathcal{U}(32,64)$ and compute the loss on the final prediction after T steps. We minimize standard softmax cross-entropy over the output logits. No additional class weighting or auxiliary losses are used.

The segmentation performance of the NCA model across all benchmarks is summarized in Table~\ref{tab:segmentation_results}. We report Pixel Accuracy, Mean IoU, and Dice Coefficient to provide a baseline for the model's efficacy prior to uncertainty evaluation. 

While the NCA achieves high performance on datasets like DSB2018, the lower mIoU scores on Kvasir-SEG and ClinicDB reflect the inherent difficulty of these tasks, providing a diverse range of error regimes for evaluating uncertainty methods.

\subsection{Baselines for Uncertainty Quantification}
The proposed \textit{resilience} method is compared against uncertainty signals derived from NCA outputs and trajectories. Each method assigns a scalar uncertainty score $u$ per test image which is used to rank predictions from most to least confident for selective prediction and failure detection (See Table \ref{tab:results}). For map-based methods, a pixel-wise uncertainty map $U(x)$ is computed and aggregated over a thin boundary band $B(m)$ around the predicted mask $m$ (dilation minus erosion). Ranking uses the boundary mean
$
u=\frac{1}{|B|}\sum_{x\in B(m)}U(x)
$
while the 95th percentile (p95) within $B(m)$ is used only as an additional reporting statistic and not for ranking.

\paragraph{Single-pass entropy ("\textit{single}").}
A single rollout to $T$ yields softmax probabilities; uncertainty is the per-pixel entropy map $U(x)=-\sum_c p_c(x)\log p_c(x)$.

\paragraph{Stopping-time disagreement ("\textit{stoptime}").}
Masks from intermediate rollouts with $T_k\sim\mathcal{U}(T_{\min},T_{\max})$ are compared to the final mask; $U(x)$ is the fraction of steps where $m^{(T_k)}(x)\neq m^{(T_{\max})}(x)$.

\paragraph{Late-stage drift ("\textit{stability}").}
Over the last $W$ steps, $U(x)$ is the mean absolute change in foreground probability, $U(x) = \frac{1}{W}\sum |p_t(x)-p_{t-1}(x)|$.

\paragraph{Binary flicker ("\textit{flicker}").}
Foreground probabilities are thresholded ($\tau{=}0.5$) and $U(x)$ is the flip frequency over a window of $W$ steps.

\paragraph{Test-time augmentation ("\textit{TTA}").}
Deterministic rotations/flips are applied to the input, predictions are inverse-warped and averaged; uncertainty is taken as the entropy of the averaged probability map.

\begin{table}[t]
\centering
\footnotesize
\setlength{\tabcolsep}{12pt}
\renewcommand{\arraystretch}{1.15}
\caption{Segmentation performance across five medical benchmarks. Mean and standard deviation is reported for 3 runs.}
\label{tab:segmentation_results}
\begin{tabular}{l c c c}
\toprule
\textbf{Dataset} &
\textbf{Pixel Acc.} &
\textbf{Mean IoU} &
\textbf{Dice} \\
\midrule
ClinicDB  & 0.91$\pm$0.01 & 0.58$\pm$0.01 & 0.74$\pm$0.01 \\
DSB 2018   & 0.97$\pm$0.01 & 0.89$\pm$0.01 & 0.94$\pm$0.01 \\
ISIC 2017  & 0.86$\pm$0.01 & 0.67$\pm$0.01 & 0.80$\pm$0.01 \\
Kvasir-SEG & 0.83$\pm$0.01 & 0.43$\pm$0.01 & 0.60$\pm$0.01 \\
NuInsSeg  & 0.89$\pm$0.01 & 0.62$\pm$0.04 & 0.77$\pm$0.03 \\
\bottomrule
\end{tabular}
\end{table}

\begin{table*}[h!]
\centering
\footnotesize
\setlength{\tabcolsep}{3pt}
\caption{Quantitative evaluation of uncertainty measures across datasets. Selective prediction is reported via $\Delta$Dice@90 (percentage-point Dice gain when retaining the 90\% most confident images) and AURC, and failure detection via AUROC and AUPRC. Values are mean $\pm$ std over 3 runs.}
\label{tab:results}
\begin{tabular}{l|l| c c c c}
\toprule
\textbf{Dataset} & \textbf{Method} & \textbf{$\Delta$Dice@90 $\uparrow$}  & \textbf{AURC $\downarrow$} & \textbf{AUROC $\uparrow$} & \textbf{AUPRC $\uparrow$}\\
\midrule

\multirow{6}{*}{ClinicDB}
& single                      &  3.76\% & 0.798$\pm$0.033 & 0.348$\pm$0.057 & 0.199$\pm$0.044 \\
& stability                   & -5.92\% & 0.835$\pm$0.020 & 0.437$\pm$0.078 & 0.272$\pm$0.034 \\
& stoptime                    & -7.17\% & 0.821$\pm$0.010 & 0.477$\pm$0.037 & 0.279$\pm$0.026 \\
& flicker                     & -1.17\% & 0.738$\pm$0.021 & 0.517$\pm$0.091 & 0.245$\pm$0.040 \\
& TTA                         & -0.08\% & 0.779$\pm$0.027 & 0.392$\pm$0.069 & 0.248$\pm$0.071 \\
& resilience                  &  \textbf{8.13\%} & \textbf{0.719$\pm$0.012} & \textbf{0.641$\pm$0.014} & \textbf{0.443$\pm$0.056} \\
\midrule

\multirow{6}{*}{DSB 2018}
& single                      & 3.57\% & 0.111$\pm$0.004 & 0.887$\pm$0.037 & 0.703$\pm$0.095 \\
& stability                   & 4.18\% & 0.106$\pm$0.002 & 0.948$\pm$0.029 & 0.819$\pm$0.101 \\
& stoptime                    &\textbf{4.40\%} & 0.114$\pm$0.002 & 0.921$\pm$0.029 & 0.794$\pm$0.057 \\
& flicker                     & 4.00\% & 0.109$\pm$0.001 & 0.937$\pm$0.016 & 0.824$\pm$0.037 \\
& TTA                         & 4.04\% & 0.111$\pm$0.004 & 0.873$\pm$0.048 & 0.686$\pm$0.116 \\
& resilience                  & 4.06\% & \textbf{0.103$\pm$0.003} & \textbf{0.972$\pm$0.009} & \textbf{0.914$\pm$0.034} \\
\midrule

\multirow{6}{*}{ISIC 2017}
& single                      & 1.27\% & 0.335$\pm$0.016 & 0.534$\pm$0.008 & 0.230$\pm$0.011 \\
& stability                   & 1.22\% & 0.280$\pm$0.023 & 0.623$\pm$0.025 & 0.271$\pm$0.021 \\
& stoptime                    & 1.59\% & 0.293$\pm$0.014 & 0.592$\pm$0.013 & 0.262$\pm$0.014 \\
& flicker                     & 3.39\% & 0.249$\pm$0.010 & 0.708$\pm$0.017 & 0.336$\pm$0.002 \\
& TTA                         & 4.47\% & 0.228$\pm$0.008 & 0.758$\pm$0.015 & 0.462$\pm$0.026 \\
& resilience                  & \textbf{7.61\%} & \textbf{0.176$\pm$0.003} & \textbf{0.910$\pm$0.007} & \textbf{0.755$\pm$0.004} \\
\midrule

\multirow{6}{*}{Kvasir-SEG}
& single                      & -1.73\% & 0.886$\pm$0.074 & 0.476$\pm$0.155 & 0.229$\pm$0.043 \\
& stability                   & 7.46\% & 0.873$\pm$0.015 & 0.513$\pm$0.203 & 0.276$\pm$0.097 \\
& stoptime                    & 5.45\% & 0.817$\pm$0.025 & 0.519$\pm$0.091 & 0.334$\pm$0.094 \\
& flicker                     & 0.01\% & 0.873$\pm$0.058 & 0.538$\pm$0.135 & 0.325$\pm$0.116 \\
& TTA                         & 2.08\% & \textbf{0.794$\pm$0.059} & 0.616$\pm$0.159 & 0.344$\pm$0.102 \\
& resilience                  & \textbf{9.26\%} & 0.816$\pm$0.043 & \textbf{0.718$\pm$0.105} & \textbf{0.500$\pm$0.176} \\
\midrule

\multirow{6}{*}{NuInsSeg}
& single                      & 0.55\% & 0.502$\pm$0.072 & 0.436$\pm$0.102 & 0.193$\pm$0.033 \\
& stability                   & -0.81\% & 0.546$\pm$0.082 & 0.492$\pm$0.280 & 0.369$\pm$0.255 \\
& stoptime                    & -2.23\% & 0.533$\pm$0.069 & 0.465$\pm$0.189 & 0.289$\pm$0.112 \\
& flicker                     & 1.09\% & \textbf{0.391$\pm$0.022} & \textbf{0.857$\pm$0.062} & 0.587$\pm$0.123 \\
& TTA                         & 1.21\% & 0.472$\pm$0.052 & 0.503$\pm$0.075 & 0.227$\pm$0.051 \\
& resilience                  & \textbf{7.63\%} & 0.394$\pm$0.006 & 0.797$\pm$0.150 & \textbf{0.688$\pm$0.208} \\
\bottomrule
\end{tabular}
\end{table*}

\subsection{Uncertainty Measures}
We evaluate uncertainty using selective prediction and failure detection. Each method produces an image-level uncertainty score $u$; images are ranked from most confident (low $u$) to least confident (high $u$). For selective prediction, we report $\Delta\text{Dice@90}$, defined as $100\times(\text{Dice@90}-\text{Dice})$, i.e., the percentage-point change in Dice when keeping only the 90\% most confident images compared to using all images, and AURC, the area under the risk–coverage curve with $\text{risk} =1-\text{Dice}$ on the retained set (lower is better). For failure detection, we treat $u$ as a score for identifying low-quality segmentations and report AUROC and AUPRC, which measure how well uncertainty separates poor predictions from good ones. AUPRC is most informative under class imbalance.
\begin{figure}
\centering
\includegraphics[width=\textwidth,page=3,trim=0cm 8cm 0cm 0cm,clip]{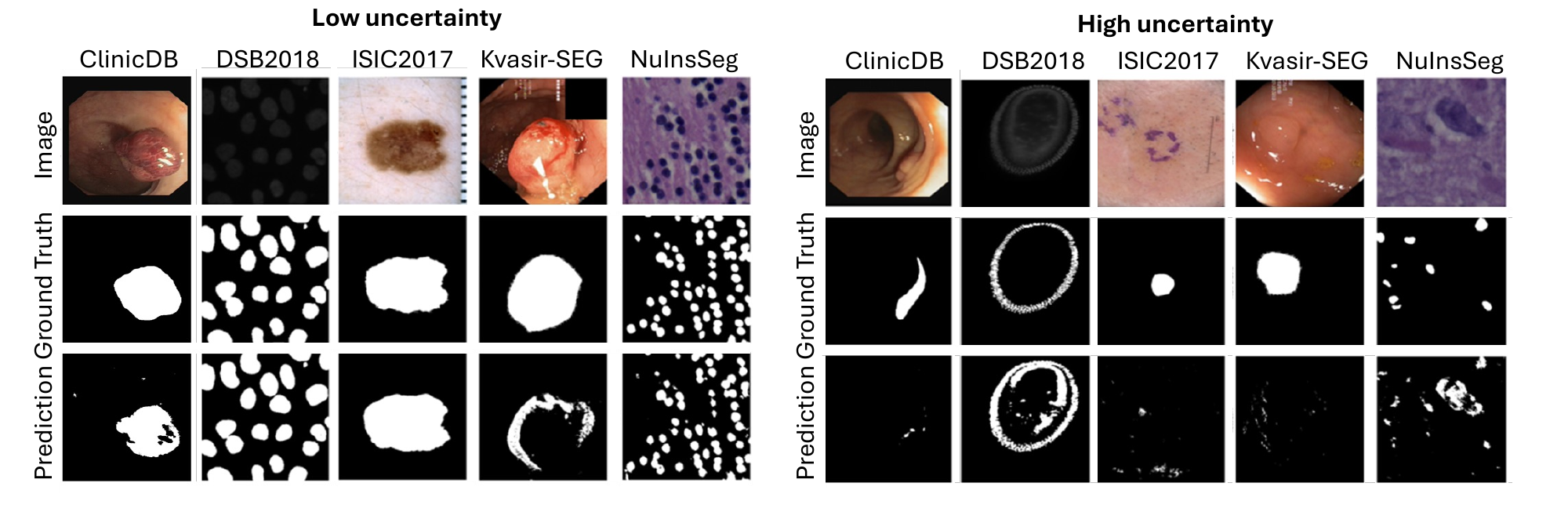}
\caption{Our \textit{resilience} method for uncertainty estimation highlights confident and failed segmentations. For each dataset, one low- and one high-uncertainty example are shown; low uncertainty corresponds to more accurate masks, while high uncertainty captures failures.}
\label{fig:qualitativeresults}
\end{figure}
\subsection{Results and Discussion}

Table~\ref{tab:results} reports selective prediction and failure-detection performance across datasets. Overall, \textit{resilience} provides the most consistent uncertainty signal, yielding strong $\Delta$Dice@90 gains and competitive failure detection. Method preferences are dataset-dependent in a few cases: \textit{stoptime} attains the best $\Delta$Dice@90 on DSB 2018, \textit{TTA} gives the lowest AURC on Kvasir-SEG, and \textit{flicker} performs best on NuInsSeg in AURC/AUROC, suggesting complementary uncertainty cues across datasets.

Figure~\ref{fig:qualitativeresults} shows typical low and high uncertainty examples. Low uncertainty cases exhibit clearer structure and more stable masks, whereas high uncertainty cases tend to involve ambiguous boundaries, low contrast, artifacts, or atypical appearance. On challenging datasets (e.g., Kvasir-SEG), even the most confident predictions may be imperfect, reflecting the difficulty of the problem rather than a failure of the uncertainty estimate.

The overall pattern indicates that probing the stability of NCA dynamics is a reliable proxy for segmentation trustworthiness: \textit{resilience} is a strong default across datasets and metrics, particularly when selective prediction at high coverage is the target. At the same time, the strongest competing baselines vary by dataset, indicating that different uncertainty families capture different failure modes. For example, \textit{stoptime} reflects sensitivity to rollout length, \textit{TTA} reflects appearance sensitivity under input transformations, and \textit{flicker} captures temporal oscillations in the predicted mask. These observations motivate the conclusion that \textit{resilience} can be used as a robust general-purpose uncertainty signal for NCA models, while hybrid or dataset-adaptive combinations may be beneficial when prioritizing a specific operating point or metric, such as minimizing AURC versus maximizing $\Delta$Dice@90.

\section{Conclusion}
We introduced \textit{resilience}, a highly efficient training-free uncertainty estimation method tailored to NCA that exploits the model’s iterative dynamics by testing whether a final prediction returns to the same solution after small state perturbations. Across five diverse biomedical segmentation benchmarks, \textit{resilience} produced a consistently strong reliability signal: it improved selective prediction (higher $\Delta$Dice@90, lower AURC) and achieved competitive failure detection (AUROC / AUPRC), while remaining lightweight and easy to integrate with existing NCA pipelines. Qualitative results show that high uncertainty concentrates on inherently challenging cases. 
Furthermore, \textit{resilience} is readout-agnostic and only probes the stability of the underlying NCA dynamics, making it directly applicable to other NCA tasks beyond segmentation.
Overall, these findings highlight that stability-driven uncertainty can serve as a simple, model-agnostic reliability layer for NCA, helping bridge the gap between label-efficient methods and clinical practicality by making errors more predictable, enabling selective use of outputs, and focusing expert attention where it matters most.

\begin{credits}
\subsubsection{\ackname} C.M. acknowledges funding from the European Research Council (ERC)
under the European Union's Horizon 2020 research and innovation program (Grant Agreement No. 866411 \& 101113551 \& 101213822) and support from the Hightech Agenda Bayern.

\subsubsection{\discintname}
The authors have no competing interests to declare that are relevant to the content of this article.
\end{credits}

%
%

\bibliographystyle{splncs04}
\bibliography{references}
\end{document}